\documentclass[aps,prc,superscriptaddress,nofootinbib,twocolumn,showpacs]{revtex4-1}
\usepackage{amsmath}
\usepackage{graphicx}

\begin{document}

\title{Magnetized strangelets at finite temperature}

\author{R. Gonz\'{a}lez Felipe}
\email{gonzalez@cftp.ist.utl.pt}
\affiliation{Instituto Superior de Engenharia de Lisboa,\\
Rua Conselheiro Em\'{\i}dio Navarro, 1959-007 Lisboa, Portugal}
\affiliation{Centro de F\'{\i}sica Te\'{o}rica de Part\'{\i}culas, Instituto Superior T\'{e}cnico,\\ Universidade T\'{e}cnica de Lisboa,
Avenida Rovisco Pais, 1049-001 Lisboa, Portugal}

\author{E. L\'{o}pez Fune}
\email{elopezf@icimaf.cu}
\affiliation{Instituto de Cibern\'{e}tica, Matem\'{a}tica y F\'{\i}sica (ICIMAF), \\
Calle E esq 15 No. 309 Vedado, Havana, 10400, Cuba}

\author{D. Manreza Paret}
\email{dmanreza@fisica.uh.cu}
\affiliation{Universidad de la Habana, Facultad de F\'{\i}sica,\\ San L\'{a}zaro y L, Havana, 10400, Cuba}

\author{A. P\'{e}rez Mart\'{\i}nez}
\email{aurora@icimaf.cu}
\affiliation{Instituto de Cibern\'{e}tica, Matem\'{a}tica y F\'{\i}sica (ICIMAF), \\
Calle E esq 15 No. 309 Vedado, Havana, 10400, Cuba}

\begin{abstract}
The main properties of strangelets, namely, their energy per baryon, radius and electric charge, are studied in the unpaired magnetized strange quark matter (MSQM) and paired magnetized color flavor locked (MCFL) phases. Temperature effects are taken into account in order to study their stability compared to the $^{56}$Fe isotope and nonmagnetized strangelets within the framework of the MIT bag model. We conclude that the presence of a magnetic field tends to stabilize more the strangelets, even when temperature is considered. It is also shown that MCFL strangelets are more stable than ordinary MSQM strangelets for typical gap values of the order of $\mathcal{O}(100)$~MeV. A distinctive feature in the detection of strangelets either in cosmic rays or in heavy-ion collider experiments could be their electric charge. We find that the electric charge is modified in the presence of the magnetic field, leading to higher (lower) charge values for MSQM (MCFL) strangelets, when compared to the nonmagnetized case.
\end{abstract}



\maketitle

\section{Introduction}

The idea of strange quark matter (SQM) as the ground state of nuclear matter, proposed by Bodmer-Witten-Terazawa~\cite{Bodmer:1971we,Witten:1984rs,Terazawa}, and enforced by Farhi and Jaffe~\cite{Farhi:1984qu}, has continued alive during the last three decades. Even more exciting are the results about the theoretical possibility to have the SQM in a neutral paired phase - the so-called color flavor locked (CFL) phase~\cite{Bailin:1983bm,Alford:2001zr}, where all quark flavors and colors are paired. The true ground state of matter would be in this paired phase~\cite{Alford:2001zr}, which is expected to occur at high densities and low temperatures.

If the SQM or CFL phases are the fundamental states of matter, one could expect them to be present inside compact objects, such as quark or hybrid stars~\cite{Ivanenko:1965dg,Itoh:1970uw}, or in the even more exotic small quark lumps named strangelets. The formation of these lumps could be caused by the collision of compact stars, supernova explosions or primary cosmic rays from strange stars.  They could even be produced in ultra-relativistic heavy-ion colliders, such as the RHIC~\cite{RHIC}, and in the Large Hadron Collider (LHC) experiments, confirming with their signatures the formation of the quark-gluon plasma (QGP). Furthermore, experiments pointing to cosmic rays are on the search for strangelets~\cite{Klingenberg:2001qs,Finch:2006pq}.

In early works~\cite{Farhi:1984qu,Berger:1986ps,Gilson:1993zs,Heiselberg:1993dc}, and in more recent papers~\cite{Chao:1995bk,Madsen:1998uh,Wen:2005uf}, several studies on SQM in the framework of the phenomenological MIT bag model, and on the properties of strange stars and strangelets, have been tackled. In particular, they have analyzed the finite size effects of strangelets and discussed their stability at $T=0$ with an energy per baryon lower than the corresponding one of the iron isotope $^{56}$Fe, the most stable nucleus known in Nature. At $T\neq 0$, strangelets would be in metastable states, but even if they survive $0.01$ milliseconds, their signatures would be very important for relativistic heavy-ion collider experiments~\cite{Madsen:1998uh}.

The above studies also suggest that the radii $R$ of spherical strangelets have a dependence with the baryon number $A$ of the form $R=r_{0}A^{1/3}$~\cite{Berger:1986ps}, where $r_{0}$ is the reduced strangelet radius, which is approximately equal to 1.12~fm for the nucleus in the atmosphere. These estimates could be of great interest in the search of strangelets upcoming from cosmic rays and reaching the Earth's atmosphere. For example, the mean free path of strangelets in the atmosphere strongly depends on their radii~\cite{Wen:2007ug}.

An interesting characteristic of strangelets arises from their relatively low baryon numbers. For $A \lesssim 10^{7}$, electrons (positrons) cannot be trapped and mixed with quarks inside strangelets due to their small Compton's wavelength compared to the radius of strangelets~\cite{Madsen:1998uh}. This implies that electric charge neutrality is in general no longer ensured, opposite to what happens in strange stars. A net nonzero electric charge for strangelets is an important parameter to be measured in order to clarify whether their existence and stability are possible in the Universe, and to establish definite conclusions on the QGP formation. The sign of the electric charge will also give us information about the particle composition, and some authors even attempt to propose possible epic cataclysmic scenarios depending on it: positively charged strangelets would repel ordinary nuclei, but negatively charged ones would attract them, turning into strange nuclei, so that the normal matter we know would change, leading to a great disaster~\cite{Madsen:1998uh}.

Most of the experiments searching for strangelets are sensitive only to low baryon number values ($A\lesssim 10^{4}$). Thus, it is important to know their theoretical properties. Since QCD equations cannot be solved analytically, nor by perturbative methods, we shall use in our study the bag model, which treats quarks (anti-quarks) as a noninteracting particles confined into a spherical well. The latter may be thought as an external parameter that realizes the color confinement of quarks and gluons. Several results about the stability, radius and electric charge of strangelets can be obtained from the liquid drop model (LDM)~\cite{Farhi:1984qu,Madsen:1998uh}, widely used in hadron physics. In this case, strangelets are described as non-interacting Fermi and gluon gases with a modified state density, using the multiple reflection expansion method (MREM) developed by Balian and Bloch~\cite{Balian:1970fw} (see also Refs.~\cite{Chao:1995bk,Madsen:2001fu,Paulucci:2008jd,Mustafa:1997ga,Wen:2005uf}).

In astrophysical scenarios as well as in heavy ion colliders, the magnetic field is an important physical quantity. Pulsars, magnetars, neutron stars, and the emission of intense sources of X-rays could be associated to compact objects with an intense magnetic field at the surface around $10^{13}-10^{15}$~G  ~\cite{Duncan:1992hi,Kouveliotou:1998ze}. In the interior of these compact objects, the equipartition theorem predicts magnetic fields of the order of $10^{19}-10^{20}$~G~\cite{Ferrer:2010wz}. Recent studies also indicate that very intense magnetic fields, reaching values of~$10^{19}-10^{20}$~G, might be generated in heavy ion colliders~\cite{Fukushima:2008xe}.

The purpose of the present paper is to study the main properties of strangelets at finite temperature in the presence of a strong magnetic field. Our analysis is done within the framework of the bag model, and using the LDM approximation~\cite{Chao:1995bk,Madsen:1998uh,Paulucci:2008jd,Wen:2005uf}. By including temperature effects, our results can be applicable to astrophysical as well as heavy-ion collider environments. In order to take account of the finite size effects, we also consider nonzero quark masses and the effects of the magnetic field and temperature on the surface and curvature energy density corrections to the quark and gluon gas. We then aim at quantifying the relevant parameters in the characterization of strangelets and their stability.

The paper is organized as follows. In Sec.~\ref{sec2} we present the general thermodynamical expressions to describe strangelets at finite density and temperature in the presence of a strong magnetic field. We also discuss the criterium of mechanical equilibrium based on the minimization of the free energy, and briefly address the issue of the Debye screening of the strangelet electric charge. This effect, although not relevant for paired MCFL (CFL) strangelets, it is of major importance in determining the total charge of unpaired MSQM (SQM) strangelets. In Sec.~\ref{sec3} we present the numerical results for the main properties of MSQM and MCFL strangelets, namely, their energy per baryon, radius and charge. We also study how the critical baryon number, below which strangelets are unstable, depends on several input parameters. Finally, our concluding remarks are given in Sec.~\ref{sec4}.

\section{Magnetized strangelets at $T \neq 0$ and $\mu \neq 0$}
\label{sec2}

The phenomenological MIT bag model has been widely used to describe strangelets with low baryon numbers~\cite{Farhi:1984qu,Berger:1986ps,Gilson:1993zs,Heiselberg:1993dc,Madsen:1998uh}. It has been shown that surface and curvature phenomena are indeed important and should be considered. In particular, surface and curvature corrections to the thermodynamical potential, energy and particle density play an important role on their stability at low baryon number~\cite{Madsen:1998uh}. At high baryon number, these quantities approach asymptotically the bulk limit, in which the surface and curvature properties are negligible. This leads to an unbound phase of SQM, which has been widely studied in order to get a more comprehensive understanding of the interior of neutron and hybrid stars.

To consider the effects of the magnetic field on the thermodynamical properties of strangelets, we first need to know the energy spectrum of the particles involved in the system. We assume that our system is under the action of a constant and homogeneous magnetic field $\mathcal{B}$, guided in the $z$-direction. The energy levels are quantized by Landau levels in the plane perpendicular to the field so that the energy spectrum is~\cite{Felipe:2007vb}
\begin{equation}\label{espectroquarks}
E_{p,f}^{\nu,\eta}=\sqrt{p_{z}^{2}+p_{\perp f}^{\;2}+m_{f}^{2}}\,,
\end{equation}
where
\begin{equation}\label{pperp}
p_{\perp f}=m_{f}\sqrt{\dfrac{\mathcal{B}}{\mathcal{B}_{f}^{c}}\bigl(2n-\eta+1\bigr)}\,,\quad
\mathcal{B}_{f}^{c}=\dfrac{m_{f}^{2}}{q_{f}},
\end{equation}
$m_{f}$ are the quark masses, $f=(u,d,s)$, the quantity $n$ indexes the Landau level, $\eta=\pm 1$ are the quark spin projections onto the magnetic field direction, $\mathcal{B}_{f}^{c}$ are the critical magnetic fields and $q_{f}$ denote the quark electric charges.

In the absence of a magnetic field, strangelets can be considered as spherical drops, characterized by their volume $V=4\pi R^{3}/3$, surface area $S=4\pi R^{2}$ and extrinsic curvature $C=8\pi R,$ where $R$ is the radius of the sphere. On the other hand, in the presence of strong magnetic fields ($\gg 10^{18}$~G), strangelets are expected to loose their sphericity due to the breaking of the spatial rotational symmetry, which also leads to the splitting of the parallel and perpendicular (to the field direction) pressure components~\cite{Felipe:2007vb,Felipe:2008cm,Ferrer:2010wz}. Nevertheless, in our study we shall assume a spherical shape for strangelets, which turns out to be a good approximation for $\mathcal{B} \lesssim 5\times10^{18}$~G. Deformed nonmagnetized strangelets have been considered, e.g., in Ref.~\cite{Mustafa:1997ga}, showing that spherical strangelets exhibit an energy per baryon lower than the deformed ones, and thus being more stable.

When dealing with massive $u$, $d$, and $s$ quarks, as well as with a gluon gas, the general expression for the thermodynamical potential can be written as
\begin{align}
\Omega\;\;\;&=\Omega_{g}+\Omega_{q\overline{q}},\\
\Omega_{g}\;\,&=\Omega_{g,v}\,V+\Omega_{g,c}\,C,\\
\Omega_{q\overline{q}}&=B_\text{bag}\,V+\sum_{f}\left[\Omega_{f,v}\,V+\Omega_{f,s}\,S+\Omega_{f,c}\,C\right].\label{OTotal}\,
\end{align}
The quantity $\Omega_{g}$ takes into account the gluon contribution to the thermodynamical potential, while $\Omega_{q\overline{q}}$ corresponds to the quark $q$ (anti-quark $\overline{q}$) contribution. Gluons, being massless particles, do not contribute to the surface corrections~ \cite{Farhi:1984qu,Madsen:1998uh,Madsen:2001fu}. Up to one-loop approximation, the bulk and curvature gluon terms can be analytically obtained and read
\begin{align}\label{Ogluon}
\Omega_{g,v}(T)=-\frac{d_{g}\pi^{2}}{90}\,T^{4},\quad \Omega_{g,c}(T)=\frac{d_{g}}{36}\,T^{2},
\end{align}
where $d_{g}=16$ is the gluon statistical weight.

The term $B_\text{bag}\,V$ accounts for the bag energy, with $B_\text{bag}$ being interpreted as a vacuum pressure that realizes the color confinement of quarks. The sum in the last three terms of Eq.~\eqref{OTotal} is over all quark flavors.

In the one-loop approximation, the bulk statistical contribution to the quark thermodynamical potential given in Eq.~\eqref{OTotal} reads
\begin{equation}\label{OVolumen}
\Omega_{f,v}(\mu_{f},T)=-\dfrac{d_{f}T}{(2\pi)^{3}}\int \ln
\bigl(f_{p}^{+}f_{p}^{-}\bigr)\,d^{3}p,
\end{equation}
where
\begin{equation}\label{FermiDiracDistrib}
f_{p}^{\pm}=1+e^{-(E_{p,f}\mp\mu_{f})/T}
\end{equation}
represents the distribution function of particles ($f_{p}^{+}$) and anti-particles ($f_{p}^{-}$) at the temperature $T$; $d_{f}=3$ is the quark statistical weight and $\mu_{f}$ is the chemical potential per gas flavor.

The bulk thermodynamical properties of SQM~\cite{Hansel,Weber} and MSQM~\cite{Felipe:2007vb,Felipe:2008cm} have been studied in the astrophysical context, which is of great importance in describing the equation of state of the quark gas, as well as the mass-radius relation for quark stars.

The surface correction effects for quarks given in Eq.~\eqref{OTotal} can be written as~\cite{Berger:1989tw}
\begin{align}\label{OSuperficie}
\begin{split}
\Omega_{f,s}&=\dfrac{d_{f}T}{16\pi^{3}}\int G_{s}\ln
\left(f_{p}^{+}f_{p}^{-}\right)\dfrac{d^{3}p}{|\vec{p}|},\\
G_{s}\;\,&=\arctan \left(m_{f}/|\vec{p}|\right).
\end{split}
\end{align}
The factor $G_{s}$ takes into account the modification of the state density within the MREM. For massless particles, it vanishes even at finite temperature. For massive particles, their densities on the surface remains always negative, i.e., particles shy away from the surface as a consequence of the boundary condition in the bag model. We also note that, in the limit $T=0$, and neglecting the $u$ and $d$ quark masses, Eq.~\eqref{OSuperficie} leads to the expression obtained by Berger and Jaffe~\cite{Berger:1986ps}.

In an analogous fashion, the curvature energy density correction to the thermodynamical potential of quarks can be written as~\cite{Madsen:1994vp,Madsen:1998uh}:
\begin{align}\label{OCurvatura}
\begin{split}
\Omega_{f,c}&=-\dfrac{d_{f}T}{48\pi^{3}}\int G_{c}\ln
\left(f_{p}^{+}f_{p}^{-}\right)\dfrac{d^{3}p}{|\vec{p}|^{2}}\,,\\
G_{c}\;\,&=1-\dfrac{3}{2}\dfrac{|\vec{p}|}{m_{f}}\arctan\left(m_{f}/|\vec{p}|\right).
\end{split}
\end{align}
In this case, the factor $G_{c}$ has been derived~\cite{Madsen:1998uh} to fit the shell model.

Since the anisotropy in the spectrum, imposed by the presence of a magnetic field, quantizes the energy states of every particle, the integration over $dp_{x}dp_{y}$ in Eqs.~\eqref{OVolumen}, \eqref{OSuperficie} and \eqref{OCurvatura} are to be replaced by the rule
\begin{equation}
\int_{-\infty}^{+\infty}\int_{-\infty}^{+\infty} dp_{x}\,dp_{y} \rightarrow 2\pi q_{f} \mathcal{B} \sum_{\eta=\pm1} \sum_{n=0}^{n_{\mbox{\tiny{max}}}^{f}},
\end{equation}
where the sum over Landau levels is limited up to
\begin{equation}\label{Landaulevels}
n_{\mbox{\tiny{max}}}^{f}=I\left[\dfrac{x_{f}^{2}-1}{2\mathcal{B}/\mathcal{B}_{f}^{c}}\right],
\end{equation}
due to the fact that the Fermi momenta must be real-valued quantities~\cite{Felipe:2007vb}. Here we have denoted
by $x_{f}=\mu_{f}/m_{f}$ the dimensionless quark chemical potentials and $I[x]$ the integer part of the real number $x$.

The role of the magnetic field in the quantities associated to surface and curvature phenomena has been studied in Ref.~\cite{Chakrabarty:1996te}. Due to the presence of a magnetic field these quantities exhibit divergences for the ground-state Landau level ($n=0$). Indeed, when $p_{\perp f}=0$, the surface contribution of Eq.~\eqref{OSuperficie} diverges logarithmically as $\sum_{f} \Omega_{f,s} \sim -\mu \ln (p_{\perp u}\,p_{\perp d}\,p_{\perp s})$, while the curvature energy density given in Eq.~\eqref{OCurvatura} diverges as $\sum_{f}\Omega_{f,c} \sim
-\mu (1/p_{\perp u}+1/p_{\perp d}+1/p_{\perp s})+\mu\ln (p_{\perp u}\,p_{\perp d}\,p_{\perp s})$. Since these divergences only appear in the limit $p_{\perp f}=0$, they can be removed by introducing an infrared cutoff in $p_{\perp f}$\footnote{We note that the inclusion of the anomalous magnetic moment of quarks in the energy spectrum acts as a natural infrared cutoff for $p_{\perp f}$~\cite{Felipe:2007vb}.}. Above this cutoff, the surface and curvature terms are then numerically well-behaved quantities.

To study the stability of strangelets the free energy must be minimized. This minimum represents the most stable configuration that the quark droplet can reach. The free energy can be written as
\begin{align}\label{freeenergy}
\begin{split}
F\;\;\,&=F_{g}+F_{q\overline{q}},\\
F_{g}\;\,&=\Omega_{g,v}\,V+\Omega_{g,c}\,C,\\
F_{q\overline{q}}&=B_\text{bag}\,V+\sum_{f}\left[F_{f,v}\,V+F_{f,s}\,S+F_{f,c}\,C\right],
\end{split}
\end{align}
where
\begin{align}
\begin{split}
F_{f,v}&=\Omega_{f,v}+\mu_{f}N_{f,v},\\
F_{f,s}&=\Omega_{f,s}+\mu_{f}N_{f,s},\\
F_{f,c}&=\Omega_{f,c}+\mu_{f}N_{f,c}.
\end{split}
\end{align}
Bulk, surface and curvature corrections to the particle density are
\begin{equation}
N_{f,v}=-\frac{\partial\Omega_{f,v}}{\partial\mu_{f}},\;
N_{f,s}=-\frac{\partial\Omega_{f,s}}{\partial\mu_{f}}, \;
N_{f,c}=-\frac{\partial\Omega_{f,c}}{\partial\mu_{f}},
\end{equation}
respectively, so that the total particle density reads
\begin{align}\label{Nparticle}
N_{f}=N_{f,v}\,V+N_{f,s}\,S+N_{f,c}\,C.
\end{align}
In turn, the baryon number $A$ is
\begin{align}\label{barionicnumber}
A =\frac{1}{3}\sum_{f}N_{f}.
\end{align}

Equilibrium configurations can be found by solving the equation
\begin{align} \label{equilconf}
\left.\frac{\partial F}{\partial V}\right|_{N,T,\mathcal{B}}=0.
\end{align}
The energy of the gas is obtained from the free energy by adding the entropy contribution, i.e., $E=F+T\,S$. The entropy $S$ contains the bulk, surface and curvature contributions,
\begin{equation}
S_{f,v}=-\frac{\partial\Omega_{f,v}}{\partial T},\;
S_{f,s}=-\frac{\partial\Omega_{f,s}}{\partial T}, \;
S_{f,c}=-\frac{\partial\Omega_{f,c}}{\partial T},
\end{equation}
respectively.

As electrons (positrons) are forbidden to coexist with quarks in strangelets due to their small radii, electric charge neutrality is never ensured. This is an important feature in their detection. The electric charge, derived from the free charge distribution, is given by the expression $Z_{\text{free}} = z_{v}+z_{s}$, where
\begin{align}\label{freecharge}
z_v&=\sum_f q_f N_{f,v}\,,\\
z_s&=\sum_f q_f N_{f,s}S+\sum_f q_f N_{f,c}C.
\end{align}

For strangelets, however, the effects of the Debye screening are not negligible and should be taken into account~\cite{Heiselberg:1993dc,Endo:2005zt,Alford:2006bx}. The strangelet charge distribution will depend on the Debye screening length $\lambda_D$,
\begin{equation}
\lambda_D^{-2}= 4\pi \sum_f q_f^2 \frac{\partial N_{f,v}}{\partial \mu_f}.
\end{equation}
By solving Poisson's equation in the Thomas-Fermi approximation, the screened volumetric charge is then calculated as~\cite{Heiselberg:1993dc}
\begin{equation}\label{Zscreen}
Z_{v}=\frac{4\pi}{e} R \lambda_D^2\, z_v \left[1-\frac{\lambda_D}{R}\tanh\left(\frac{R}{\lambda_D}\right)\right],
\end{equation}
and the total electric charge becomes
\begin{equation}\label{Ztotal}
Z=Z_{v}+z_{s}.
\end{equation}
The corresponding Coulomb energy of the strangelet is given by~\cite{Heiselberg:1993dc}
\begin{align}
E_C=4\pi^2 z_v^2 \lambda_D^4 \left[2-\frac{3\lambda_D}{R}\tanh\left(\frac{R}{\lambda_D}\right) +\cosh^{-2} \left(\frac{R}{\lambda_D}\right)\right],\nonumber\\
\end{align}
which should be added to the total energy of the system.

Finally, it is worthwhile to mention that, in the study of strange stars, one can derive from Eq.~\eqref{OVolumen} all the thermodynamical bulk properties and the equation of state (EoS) of the system. In this case, the baryon number $A$ is a huge quantity ($A \sim 10^{57}$) so that the surface and curvature contributions can be safely neglected. On the other hand, to describe the mechanical equilibrium configurations of strangelets, where $A \leq 10^{7}$, the finite size effects, like the surface tension in Eq.~\eqref{OSuperficie} and the curvature energy density in Eq.~\eqref{OCurvatura}, play an important role in determining their properties. These terms are also important in the study of the quark-hadron phase transition~\cite{Horvath:1994zn,Madsen:1998uh,Chakrabarty:1996te}.

\section{Numerical results: MSQM strangelets and MCFL strangelets}
\label{sec3}

In this section we solve numerically Eq.~\eqref{equilconf}, which guarantees the mechanical equilibrium of strangelets. Our aim is to study the energy per baryon for strangelets in the MSQM and MCFL phases as a function of the baryon number $A$, bag energy density, the magnetic field $\mathcal{B}$, temperature $T$ and gap parameter. We shall compare our results with the case of zero magnetic field, and also compute the radius of strangelets and their electric charge as a function of the baryon number. First we consider strangelets formed from unpaired MSQM, and then we study those corresponding to the paired MCFL phase.

\subsection{Unpaired phase: MSQM strangelets}

Strangelets with $A\leq10^{7}$ require zero electron and positron number densities. If they are formed from SQM or MSQM, the $\beta$-equilibrium condition implies the equality of chemical potentials: $\mu_{u}=\mu_{d}=\mu_{s}$. In the presence of a magnetic field, however, the latter condition does not imply the equality of bulk particle densities $N_{f,v}$ in MSQM strangelets. Indeed, we recall that, at $T=0$,
\begin{equation}
N_{f,v}=N^0_{f}\sum_{\eta=\pm1}\sum_{n=0}^{n_{\mbox{\tiny{max}}}^{f}}
p_{F}^{f}\,, \quad
N_{f}^0=\frac{d_{f}m_{f}^3}{2\pi^2}\dfrac{\mathcal{B}}{\mathcal{B}^{c}_{f}},\label{TQf}
\end{equation}
where
\begin{equation}\label{magmass}
p_{F}^{f}= \sqrt{x_{f}^2-h_{f}^{\eta\;2}}\,,\quad
h_{f}^{\eta} =\sqrt{\frac{\mathcal{B}}{\mathcal{B}^{c}_f}\, (2n-\eta+ 1) +1}\,,
\end{equation}
correspond to the $z$-component of the magnetic Fermi momenta and the magnetic mass, respectively~\cite{Felipe:2007vb}. The presence of $h_{f}^{\eta}$ in the Fermi momenta and
$\mathcal{B}^{c}_{f}$ in the particle densities implies that, despite the chemical potential equalities, the particle densities should be different for every flavor.

Equation~\eqref{equilconf} can be written in the form
\begin{align}
\left.\dfrac{\partial F_{g}}{\partial V}\right|_{N,T,\mathcal{B}}+
\left.\dfrac{\partial F_{q\overline{q}}}{\partial
V}\right|_{N,T,\mathcal{B}}=0,\label{equilibrio}
\end{align}
with
\begin{align}\label{equilibrio1}
\left.\dfrac{\partial F_{g}}{\partial
V}\right|_{N,T,\mathcal{B}}&=\Omega_{g,v}+\dfrac{2}{R^{2}}\Omega_{g,c},\\
\left.\dfrac{\partial F_{q\overline{q}}}{\partial
V}\right|_{N,T,\mathcal{B}}&=B_\text{bag}+ \frac{\partial E_C}{\partial V}+\Omega_{v}+ \dfrac{2}{R}\Omega_{s}+\dfrac{2}{R^{2}}\Omega_{c},\label{equilibrio2}
\end{align}
and
\begin{align}\label{equilibrio3}
\Omega_{v}=\sum_{f}\Omega_{f,v},\quad \Omega_{s}=\sum_{f}\Omega_{f,s},\quad \Omega_{c}=\sum_{f}\Omega_{f,c}.
\end{align}
The above system of equations, together with Eq.~\eqref{barionicnumber}, allows us to determine the common quark chemical potential and the radius of the strangelet, as well as to evaluate other quantities such as the energy per baryon, particle densities, and electric charge.

\begin{figure}[t]
\centering
\includegraphics[width=0.5\textwidth]{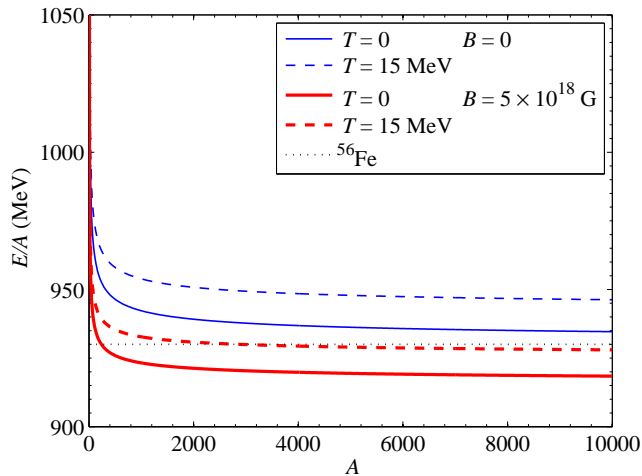}\\
\caption{(Color online) Energy per baryon for unpaired MSQM and SQM strangelets with $\mathcal{B}=5\times 10^{18}$~G. We take $B_\text{bag}=75$~MeV fm$^{-3}$ and consider $T=0, 15$~MeV.} \label{AEA}
\end{figure}

\begin{figure}[t]
\begin{center}
\includegraphics[width=0.5\textwidth]{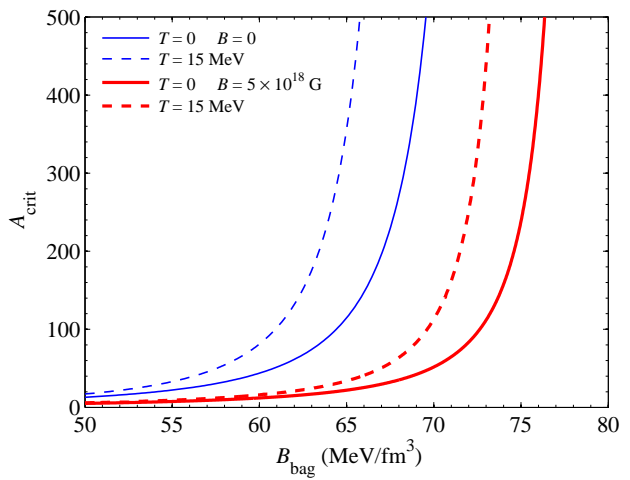}
\end{center}
\caption{(Color online) Critical baryon number $A_\text{crit}$ of MSQM strangelets for
$\mathcal{B}=5\times10^{18}$~G and $T=0, 15$~MeV, as a function of $B_\text{bag}$.
The nonmagnetized SQM case is also shown.}\label{AcritSQM}
\end{figure}

\begin{figure}[t]
\begin{center}
\includegraphics[width=0.5\textwidth]{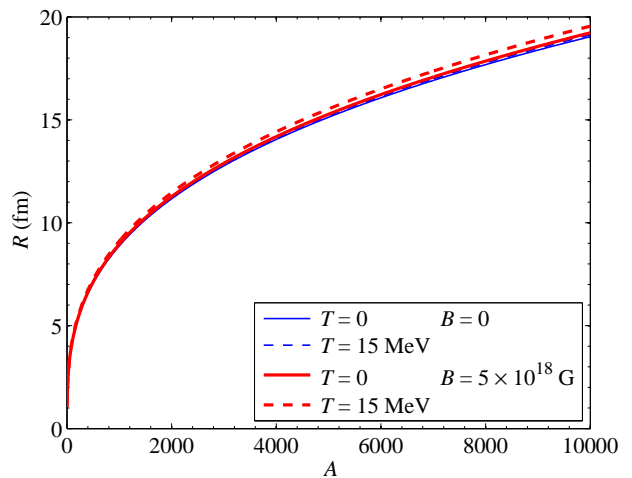}
\end{center}
\caption{(Color online) Radii of strangelets as a function of the baryon number for
$B_\text{bag}=75$~MeV fm$^{-3}$ and $T=0, 15$~MeV. Radii of MSQM
($\mathcal{B}=5\times10^{18}$~G) and SQM ($\mathcal{B}=0$) strangelets are depicted.}\label{RA}
\end{figure}

\begin{figure}[t]
\begin{center}
\includegraphics[width=0.5\textwidth]{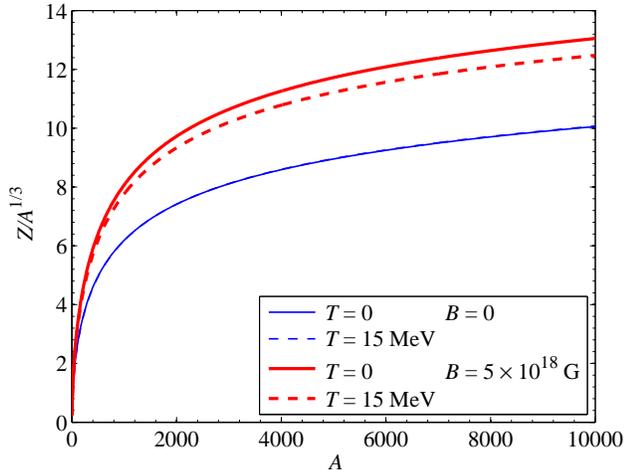}
\end{center}
\caption{(Color online) Ratio $Z/A^{1/3}$ (in units of $e$) for
$\mathcal{B}=5\times10^{18}$~G and $T=0, 15$~MeV. We take
$B_\text{bag}=75$~MeV fm$^{-3}$. The nonmagnetized SQM
strangelet electric charge is also shown.}\label{ZA}
\end{figure}

In Fig.~\ref{AEA} we present the energy per baryon for unpaired MSQM strangelets at zero and finite temperature, assuming $m_{u}=m_{d}=5$~MeV, $m_{s}=150$~MeV, $B_\text{bag}=75$~MeV fm$^{-3}$ and $\mathcal{B}=5\times10^{18}$~G. The energy per baryon of SQM ($\mathcal{B}=0$) is also shown for comparison. We observe that MSQM strangelets exhibit an $E/A$ lower than SQM strangelets. The horizontal dotted line corresponds to the iron energy per baryon. We assume $\left.\frac{E}{A}\right|_{\mathcal{B}\neq0}(^{56}\text{Fe})\simeq \left.\frac{E}{A}\right|_{\mathcal{B}=0}(^{56}\text{Fe})=930$~MeV, which is a consistent approximation in the magnetic field range $\mathcal{B} < 10^{19}$~G~\cite{Martinez:2003dz,Paulucci:2010uj}.

The presence of the magnetic field reduces $E/A$ even at $T=15$~MeV, for the bag parameter and quark masses considered. Finite lumps of magnetized quark matter with high baryon numbers could be absolutely stable at this temperature, which opens the possibility of finding strangelets from the coalescence of quark or hybrid stars where those high magnetic fields are expected to be present. Clearly, temperature effects increase $E/A$ due to the increment of the thermal energy of quarks and gluons.

For both unpaired MSQM and SQM strangelets, the energy per baryon as a function of the baryon number can be fitted with the general expression~\cite{Madsen:1998uh}
\begin{equation}\label{AEAgeneral}
\dfrac{E}{A}=E_{v}+\dfrac{E_{s}}{A^{1/3}}+\dfrac{E_{c}}{A^{2/3}},\\
\end{equation}
where the coefficients $E_{v}$, $E_{s}$ and $E_{c}$ are ascribed to bulk, surface and curvature phenomena, respectively. This fitting function shows that, for high baryon numbers, the energy per baryon approaches its bulk value, $ E/A \sim 3\mu_{f}+T S_{v}/A$. For low baryon numbers, the terms depending on $A$ show destabilizing properties due to their growth~\cite{Farhi:1984qu,Madsen:1998uh,Paulucci:2008jd}.
As expected, the bulk limit increases with temperature. Surface and curvature terms get balanced with each other to maintain the free energy at its minimum for a fixed $A$.

We also observe from Fig.~\ref{AEA} that, in the presence of a magnetic field, strangelets can be stable for a wider range of the baryon number. Furthermore, depending on the bag parameter, temperature, and magnetic field, there exists a critical value $A_\text{crit}$ with $E/A_\text{crit}=930$~MeV, such that for $A \geq A_\text{crit}$ strangelets can be in a stable phase and long lasting in Nature, but metastable when $A < A_\text{crit}$. In Fig.~\ref{AcritSQM}, the critical baryon number is plotted as a function of the bag parameter for $\mathcal{B}=0$ and $\mathcal{B}=5 \times10^{18}$~G, and $T=0, 15$~MeV. It can be seen that for sufficiently large $B_\text{bag}$, there are no stable strangelets. For $\mathcal{B}=0$, stability requires $B_\text{bag} \lesssim 70$~MeV fm$^{-3}$~at $T=0$ and $B_\text{bag} \lesssim 66$~MeV fm$^{-3}$~at $T=15$~MeV. For $\mathcal{B}=5\times 10^{18}$~G, the upper limits are $B_\text{bag} \lesssim 77$~MeV fm$^{-3}$~at $T=0$, and $B_\text{bag} \lesssim 73$~MeV fm$^{-3}$~at $T=15$~MeV. The allowed range of bag values is thus wider in the presence of the magnetic field. It becomes clear that the magnetic field plays a crucial role on the stability of strangelets. This is observed in bulk, surface, and curvature terms. On the other hand, temperature effects turn out to be less significant in comparison with those of the magnetic field on the surface and curvature terms. Notice also that, if  $E/A > m_n \simeq 939$~MeV, SQM strangelets would exhibit neutron emission~\cite{Madsen:1998uh,Berger:1986ps}. This opens the interesting possibility to detect strangelets by the emission of particles from unknown astrophysical sources.

The dependence of the radii of SQM and MSQM strangelets on the baryon number is depicted in Fig.~\ref{RA}. Due to the thermal energy of quarks and gluons, temperature effects tend to slightly increase the radii. The same behavior is observed in the presence of a magnetic field. This is a direct consequence of the decreasing behavior of the surface and curvature terms as discussed above. The energy, preserving the stability on the surface, is diminished by the magnetic field, which tends to stabilize more strangelets, but also increases their size compared to the nonmagnetized ones. As a function of the baryon number, the strangelet radius $R$ behaves as $R =r_0 A^{1/3}$, with $r_0 \simeq 0.9$~fm, for high values of $A$. Temperature and magnetic field tend to modify the values of $r_0$ in about $1\%$, being for MSQM greater than the corresponding ones of SQM, but always smaller than the nuclear value.

Finally, the dependence of the total electric charge of strangelets on the baryon number is depicted in Fig.~\ref{ZA}. The electric charge increases with $A$ and, for a fixed $A$, it also increases with the magnetic field. For small baryon numbers, $Z \propto A$, while for large $A$ the relation between $Z$ and $A$ differs from the linear one. This is caused by the screening of the bulk's electric charge for strangelets with a radius greater than the Debye length. The strangelet electric charge is distributed within a husk of a Debye length radius from the surface. The relation $Z \propto A^{1/3}$ is observed~\cite{Heiselberg:1993dc}, when the contributions of the free surface electric charge are not considered. Including these effects, as in Eq.~\eqref{Ztotal}, we obtain the asymptotic behavior $Z \propto A^{2/3}$.

\subsection{Paired phase: MCFL strangelets}

\begin{figure}[t]
\begin{center}
\includegraphics[width=0.5\textwidth]{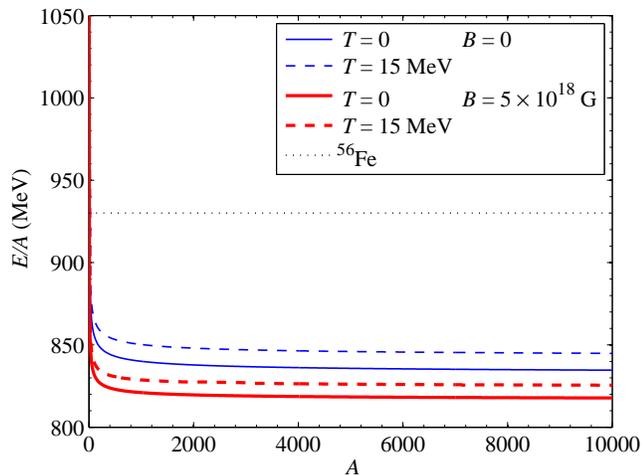}
\end{center}
\caption{\label{AEACFL}(Color online) Energy per baryon for CFL ($\mathcal{B}=0$)
and MCFL ($\mathcal{B}=5\times 10^{18}$~G) strangelets at
$T=0,15$~MeV. We fix $B_\text{bag}=75$~MeV fm$^{-3}$ and $\Delta=100$~MeV.}
\end{figure}

\begin{figure}[t]
\begin{center}
\includegraphics[width=0.5\textwidth]{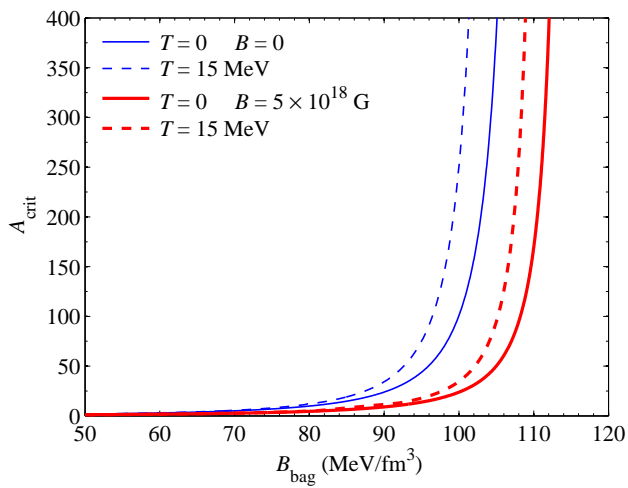}
\end{center}
\caption{(Color online) Critical baryon number $A_\text{crit}$ of CFL strangelets as a function of $B_\text{bag}$. The plots are given for $\mathcal{B}=0$ and $5\times10^{18}$~G, $T=0$ and 15~MeV, and assuming $\Delta = 100$~MeV.}\label{AcritCFL}
\end{figure}

\begin{figure}[t]
\begin{center}
\includegraphics[width=0.5\textwidth]{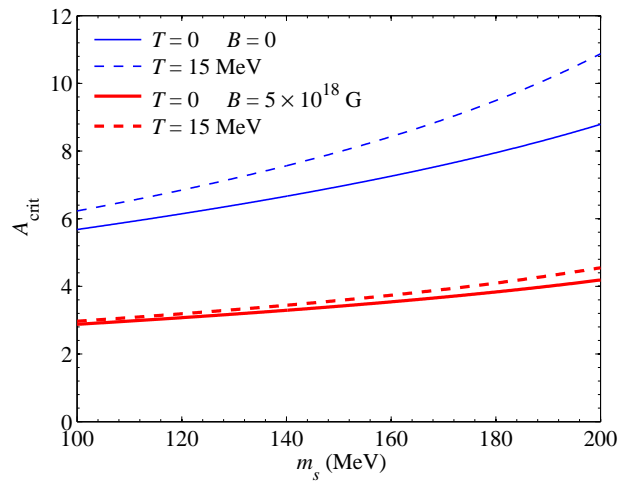}
\end{center}
\caption{(Color online) Critical baryon number $A_\text{crit}$ of CFL strangelets as a function of the strange quark mass $m_s$. The plots are given for $\mathcal{B}=0$ and $5\times10^{18}$~G, $T=0$ and 15~MeV, taking $B_\text{bag}=75$~MeV fm$^{-3}$ and $\Delta = 100$~MeV.}\label{AcritCFLms}
\end{figure}

\begin{figure}[t]
\begin{center}
\includegraphics[width=0.5\textwidth]{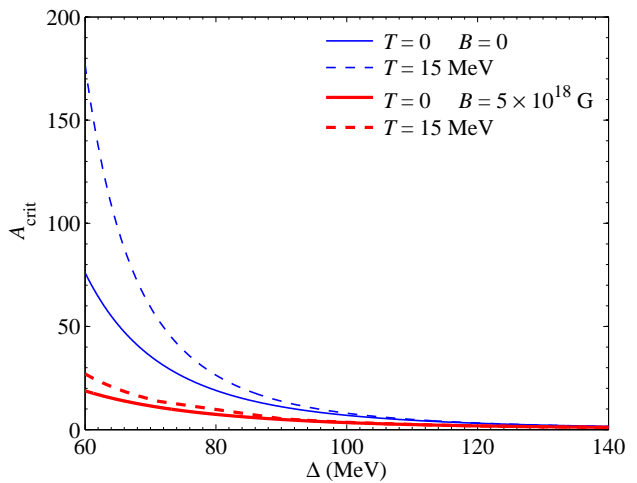}
\end{center}
\caption{(Color online) Critical baryon number $A_\text{crit}$ of CFL strangelets as a function of the gap parameter $\Delta$. The plots are given for $\mathcal{B}=0$ and $5\times10^{18}$~G, $T=0$ and 15~MeV, assuming $B_\text{bag}=75$~MeV fm$^{-3}$.}\label{AcritCFLDelta}
\end{figure}

The stability, EoS and mass-radius configurations of a quark gas in the MCFL phase have been studied in Ref.~\cite{Felipe:2010vr} within the framework of the bag model. In analogy to unpaired SQM and MSQM strangelets, we now analyze the relevant properties of strangelets composed by matter in the MCFL paired phase at finite temperature. In the limit $\mathcal{B}=0$ and $T=0$, our results agree with those obtained by Madsen in Ref.~\cite{Madsen:2001fu}.

We assume that quarks behave like noninteracting particles, but the binding energy of the diquark condensate is associated to a BCS pairing energy gap $\Delta$. Following Ref.~\cite{Felipe:2010vr}, we take a common value of the gap parameter for the predominant color pairings ($ud,us,ds$)~\cite{Fukushima:2007fc,Noronha:2007wg}. Furthermore, in order to simplify our analysis, we neglect any dependence of $\Delta$ on the magnetic field\footnote{A more accurate study would require the determination of $\Delta$ by solving the corresponding gap equations which depend on the magnetic field~\cite{Ferrer:2005vd,Paulucci:2010uj}.}. On the other hand, as our study is devoted to strangelets at finite temperature, we consider a temperature-dependent gap parameter in analogy to superconductivity studies of quark matter~\cite{Schmitt:2002sc,Alford:2007xm,Paulucci:2008jd}:
\begin{equation}\label{Delta}
\Delta=2^{-1/3} \Delta_{0}\left[1-\left(\frac{T}{T_c}\right)^{2}\right]^{1/2},
\end{equation}
where $T_{c}=2^{1/3} e^\gamma \Delta_0/\pi \simeq 0.71 \Delta_{0}$ is the critical temperature of the superconducting phase, above which the pairing of quarks is forbidden.

For the magnetized CFL phase, the bulk thermodynamical potential in Eq.~\eqref{equilibrio3} reads as
\begin{equation}\label{CFL}
\Omega_{v}=\sum_{f}\Omega_{f,v}-\frac{3\Delta^2\mu_{B}^{2}}{\pi^2},
\end{equation}
where the second term in the right hand side accounts for the cost of the free energy which is compensated by the pairing formation~\cite{Felipe:2010vr}, and $\mu_{B}=(\mu_{u}+\mu_{d}+\mu_{s})/3$ is the baryon chemical potential.

This color symmetric state requires quark bulk particle numbers to be equal to each other to minimize the free energy and enforce the color and electric charge neutrality of the CFL phase~\cite{Alford:2001zr,Alford:2002kj,Rajagopal:2000ff,Alford:2007xm,Alford:2004pf}. This requirement leads to the equations
\begin{equation}
N_{u,v}+\frac{2\Delta^2\mu_B}{\pi^2}=N_{d,v}+\frac{2\Delta^2\mu_B}{\pi^2}
=N_{s,v}+\frac{2\Delta^2\mu_B}{\pi^2}. \label{Nequality}
\end{equation}

From Eqs.~\eqref{Nequality} it is clear that the bulk electric charge for CFL and MCFL strangelets is zero, and the only contribution comes from the surface. Charge screening is thus negligible for CFL strangelets. This is in contrast with the charge behavior of SQM and MSQM strangelets, for which the bulk charge contribution is dominant.

\begin{figure}[t]
\begin{center}
\includegraphics[width=0.5\textwidth]{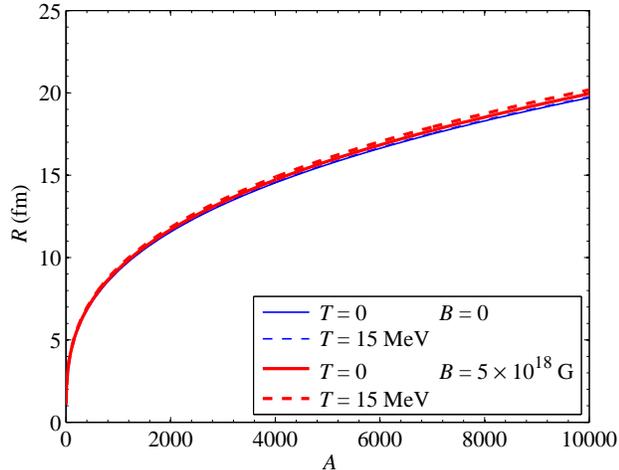}
\end{center}
\caption{(Color online) Radii of nonmagnetized and magnetized CFL strangelets for
a magnetic field value $\mathcal{B}=5\times10^{18}$~G,
$T=0,15$~MeV, $B_\text{bag}=75$~MeV fm$^{-3}$ and $\Delta=100$~MeV.}\label{RACFL}
\end{figure}

\begin{figure}[t]
\begin{center}
\includegraphics[width=0.5\textwidth]{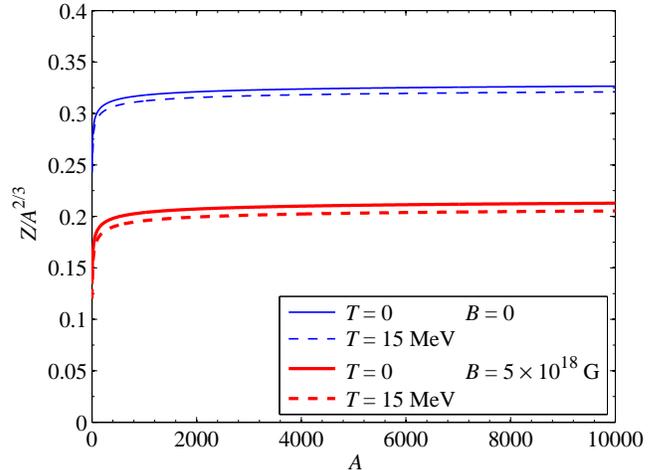}
\end{center}
\caption{(Color online) Ratio $Z/A^{2/3}$ (in units of $e$) of nonmagnetized and magnetized CFL
strangelets for $\mathcal{B}=5\times10^{18}$~G, $T=0,15$~MeV,
$B_\text{bag}=75$~MeV fm$^{-3}$ and $\Delta=100$~MeV.}\label{ZACFL}
\end{figure}

In an analogous fashion to the unpaired phase, the free energy density, which now contains the quark pairing contribution, has to be minimized by solving the equilibrium equation~\eqref{equilconf}, or equivalently, Eqs.~\eqref{equilibrio}-\eqref{equilibrio3}, with $E_C=0$ and $\Omega_{v}$ given by Eq.~\eqref{CFL}. In Fig.~\ref{AEACFL} we present the energy per baryon $E/A$ as a function of the baryon number, for nonmagnetized and magnetized CFL strangelets. The results are given for $\mathcal{B}=5 \times 10^{18}$~G, $T=0,15$~MeV, $B_\text{bag}=75$~MeV fm$^{-3}$ and $\Delta=100$~MeV. We notice that the energy per baryon, corresponding to strangelets in the paired phase, has a lower value than for unpaired strangelets. Furthermore, MCFL strangelets exhibit an $E/A$ lower than CFL strangelets.

As in the case of unpaired quark matter, surface and curvature terms are responsible for the stability of strangelets at low baryon numbers, and the energy per baryon can be fitted with a general expression of the form given in Eq.~\eqref{AEAgeneral}. At high values of $A$, surface and curvature phenomena become negligible and $E/A$ tends to its bulk value. Furthermore, bulk contribution increases with temperature, while surface and curvature get balanced with each other to maintain the free energy at a minimum. For given values of the bag and gap parameters, temperature, strange quark mass and magnetic field, the ratio $E/A$ of CFL strangelets always fixes an $A_\text{crit}$ above which the stability is guaranteed (compared to $E/A=930$~MeV). Figure~\ref{AcritCFL} shows the dependence of $A_\text{crit}$ on the bag parameter for $\mathcal{B}=0$ and $5\times10^{18}$~G, $T=0$ and 15~MeV, and taking $\Delta = 100$~MeV. Once again, for sufficiently large $B_\text{bag}$, there are no stable strangelets. For $\mathcal{B}=0$, stability requires $B_\text{bag} \lesssim 105$~MeV fm$^{-3}$~at $T=0$ and $B_\text{bag} \lesssim 102$~MeV fm$^{-3}$~at $T=15$~MeV. For $\mathcal{B}=5\times 10^{18}$~G, the upper bounds are $B_\text{bag} \lesssim 112$~MeV fm$^{-3}$~at $T=0$, and $B_\text{bag} \lesssim 110$~MeV fm$^{-3}$~at $T=15$~MeV. Therefore the presence of the magnetic field leads to a larger stability window.

To see the sensitivity of $A_\text{crit}$ to the variation of other parameters, we present in Figs.~\ref{AcritCFLms} and \ref{AcritCFLDelta} how the critical baryon number varies with the strange quark mass $m_s$ and the gap parameter $\Delta$. From Fig.~\ref{AcritCFLms} it is clear that $A_\text{crit}$ is a slowly increasing function of $m_s$. On the other hand, Fig.~\ref{AcritCFLDelta} shows that $A_\text{crit}$ is a decreasing function of the gap $\Delta$. In both cases, the effect of the temperature is to increase the required values of $A_\text{crit}$ in comparison with those at zero temperature.

The CFL and MCFL strangelet radii at $T=0,15$~MeV are shown as a function of the baryon number in Fig.~\ref{RACFL}, for fixed bag and gap parameters. It is observed the same behavior than that of the radii of the unpaired strangelets studied in the previous subsection, i.e. $R \simeq 0.9 A^{1/3}$. Furthermore, the effect of the magnetic field is to yield a larger strangelet configuration due to the contribution of surface and curvature effects. This is in contrast with magnetized compact stars, for which surface and curvature corrections are negligible, and the magnetic field yields star configurations with smaller radii. We also note that an increase of $\Delta$ would lead to higher values of the radii, as in the case of compact objects~\cite{Felipe:2010vr}.

Finally, in Fig.~\ref{ZACFL} we present the dependence of the electric charge on the baryon number for CFL and MCFL strangelets. The curves show that temperature always tends to decrease the electric charge. In the bulk, the equality of quark densities implies zero electric charge contribution. Therefore, these strangelets are characterized only by their surface electric charge distribution. As can be seen from the figure, the magnetic field also tends to decrease the surface electric charge. When $A$ is large, the behavior of $Z$ as a function of $A$ is $Z \simeq 0.3 A^{2/3}$~\cite{Madsen:2001fu} for nonmagnetized CFL strangelets, while its characteristic value is $Z \simeq 0.2 A^{2/3}$ for MCFL strangelets.

\section{Conclusions}
\label{sec4}

Strangelets from MSQM and MCFL have been studied within the MIT bag model and the LDM framework. A comparison with SQM and CFL strangelets for massive quarks has been presented. We have examined how the stability and size of quark lumps are modified by the temperature and the presence of a strong magnetic field. While temperature tends to destabilize strangelets, the magnetic field acts in the opposite direction. Any nonvanishing temperature contributes to an increase of the energy per baryon $E/A$, which measures how stable are strangelets when compared to the isotope $^{56}$Fe, or any other given pattern, such as neutrons, pions or $\Lambda$-particles. On the other hand, the magnetic field enhances their stability, both in bulk and surface. Strangelets in the MCFL and CFL phases show more stability than unpaired strangelets, as expected from the gap energy and pairing of quarks.

The stability of strangelets at zero and finite temperature strongly depends on several parameters and, particularly, on the bag constant, strange quark mass, gap parameter, and the magnetic field. There exists a critical baryon number $A_\text{crit}$ below which strangelets are metastable states, and can decay and radiate particles from their surface, even at $T=0$. The critical baryon number increases with $B_\text{bag}$, $m_s$ and $T$, but is a decreasing function of the magnetic field $\mathcal{B}$ and the gap parameter $\Delta$. These results, nevertheless, should be taken with care due to the approximate nature of the liquid drop and bag models.

The radii of strangelets were also studied in order to have an estimate of their size. As the pairing gap increases, the radii of strangelets get larger, as in the case of compact objects~\cite{Felipe:2010vr}. Magnetized strangelets turn out to have a radius bigger than nonmagnetized ones, since the magnetic field tends to relax the surface tension, thus increasing the strangelet radius.

Finally, we have also found that the charge neutrality of strangelets is never ensured and, in the presence of the magnetic field, strangelets are always positively charged. Therefore a nonvanishing electric charge may constitute the main characteristic to be searched for in experiments or in the detection of strangelets from astrophysical sources. Unlike MCFL (CFL) strangelets, which are neutral in bulk, unpaired MSQM (SQM) strangelets have a contribution to the electric charge due to the screened bulk charge. The relation $(Z,A)$ represents another characteristic that could distinguish the phase of strange quark matter present in strangelets. For large $A$, MSQM and SQM strangelet electric charge exhibits the behavior $Z \propto A^{1/3}$ (attaining $Z \propto A^{2/3}$ asymptotically), whereas for MCFL and CFL strangelets, the ratio $Z/A^{2/3}$ reaches a constant value. These could be striking properties in their detection either in cosmic rays or in heavy-ion collider experiments.

\vspace*{5mm}

\acknowledgments{The authors acknowledge the fruitful discussions with A. Gonzalez Garcia and J. Horvath,  and  their comments and suggestions. A.P.M specially thanks to Prof W. Greiner for his useful suggestions. The work of R.G.F. was supported by Funda\c{c}\~{a}o para a Ci\^{e}ncia e a Tecnologia (FCT, Portugal) through the project CFTP-FCT UNIT 777 and Grant No. CERN/FP/109305/2009, which are partially funded through POCTI (FEDER).  A.P.M, D.M.P. and E.L.F. have been supported by CITMA-Cuba under the Grant No. CB0407 and the ICTP Office of External Activities through NET-35. A.P.M. also thanks CFTP-IST in Lisbon and the Frankfurt Institute for Advance Studies (FIAS) for their hospitality and support during the scientific visits where this work was finished.}

\end{document}